\documentclass[journal]{IEEEtran}

\usepackage{graphicx}
\usepackage{multirow}
\usepackage{subfigure}

\newcommand{\replicas}{\textit{\textbf{r}}}
\newcommand{\replica}{\textit{r}}
\newcommand{\varRep}[1]{\replicas_{[#1]}}
\newcommand{\varChkRep}[2]{\replica_{[#1] #2}}
\newcommand{\varChkRepOver}[2]{\replica_{[#1] #2}^{over}}
\newcommand{\varChkRepConc}[2]{\replica_{[#1] #2}^{conc}}

\newcommand{\varAvg}[1]{\bar{r}_{[#1]}}
\newcommand{\varRepAvg}[1]{\bar{\replicas}_{[#1]}}
\newcommand{\chkRep}[1]{\replicas_{(#1)}}
\newcommand{\chkRepTil}[1]{\tilde{\replicas}_{(#1)}}

\begin{document}

\title{Divide \& Concur and Difference-Map BP Decoders for LDPC codes}

\author{Jonathan~S.~Yedidia,~\IEEEmembership{Member,~IEEE,}
  Yige~Wang,~\IEEEmembership{Member,~IEEE,} 
  Stark~C.~Draper,~\IEEEmembership{Member,~IEEE}
  \thanks{J.~S.~Yedidia and Y.~Wang are with 
Mitsubishi Electric Research Laboratories,
  Cambridge, MA 02139 (yedidia@merl.com; yigewang@merl.com).} 
  \thanks{S.~C.~Draper is with the Dept.~of Electrical and Computer
    Engineering, University of Wisconsin, Madison, WI 53706 
(sdraper@ece.wisc.edu).}}

\maketitle

\begin{abstract}
  The ``Divide and Concur'' (DC) algorithm, recently introduced by Gravel 
  and Elser, can be considered a competitor to the belief propagation (BP)
  algorithm, in that both algorithms can be applied to a wide variety of
  constraint satisfaction, 
  optimization, and probabilistic inference problems. We show that DC 
  can be interpreted as a message-passing algorithm on a constraint graph, 
  which helps make the comparison with BP more clear. The ``difference-map''
  dynamics of the DC algorithm 
  enables it to avoid ``traps'' which may be related to the ``trapping
  sets'' or ``pseudo-codewords'' 
  that plague BP decoders of low-density parity check (LDPC) 
  codes in the error-floor regime.

  We investigate two decoders for low-density parity-check (LDPC) codes based 
  on these ideas. The first decoder is based directly on DC, while the second
  decoder borrows the important ``difference-map'' concept from the DC algorithm 
  and translates it into a BP-like decoder. We show that this 
  ``difference-map belief propagation'' (DMBP) decoder has dramatically improved
  error-floor performance compared to standard BP decoders, while maintaining
  a similar computational complexity. We present simulation results for LDPC codes on
  the additive white Gaussian noise and binary symmetric channels, 
  comparing DC and DMBP decoders with other decoders based on BP, linear 
  programming, and mixed-integer linear programming.

\end{abstract}




\section{Introduction}

Properly designed low-density parity-check (LDPC) codes, decoded using
efficient message-passing belief propagation (BP) decoders, achieve
near Shannon limit performance in the so-called ``water-fall'' regime
where the signal-to-noise ratio (SNR) is near the code threshold
\cite{ModernCodingTheory}.  Unfortunately, BP decoders of LDPC codes
often suffer from ``error floors'' in the high SNR regime, which is a
significant problem for applications that have extreme reliability
requirements, including magnetic recording and fiber-optic
communication systems.

There has been considerable effort in trying to find LDPC codes and
decoders that have improved error floors while maintaining good
water-fall behavior. In general, such work can be divided into two
approaches.  The first line of attack tries to construct codes or
representations of codes that have improved error floors when
decoded using BP.  Error floors in LDPC codes using BP decoders are
usually attributed to closely related phenomena that go under the
names of ``pseudocodewords,'' ``near-codewords,'' ``trapping sets,''
``instantons,'' and ``absorbing sets''
\cite{FreyKoetterVardy}\cite{GraphCover}\cite{NearCodeword}\cite{trappingSet}\cite{Dolecek}\cite{VontobelWebpage}.
The number of these trapping sets (to choose one of these terms), and
therefore the error floor performance, can be improved by removing
short cycles in the code graph \cite{PEG}\cite{ACE}\cite{HighGirth}.
One can also consider special classes of LDPC codes with fewer
trapping sets, such as EG-LDPC
codes \cite{EGLDPC}, or generalized LDPC codes
\cite{GLDPC}\cite{DGLDPC}.

The second approach, taken herein, is to try to improve upon the
sub-optimal BP decoder.  This approach is logical because already when
he introduced regular LDPC codes, Gallager showed that they have
excellent distance properties and therefore will not have error floors
if decoded using optimal maximum-likelihood (ML) decoding
\cite{gallager}.  Building on the theory of trapping sets, Han and
Ryan propose a ``bi-mode syndrome-erasure decoder.'' This decoder can improve
error floor performance given the knowledge of dominant trapping sets
\cite{HanRyan}.  However, determining the dominant trapping sets of a particular
code can be a challenging task. Another recently introduced improved decoder is the
mixed-integer linear programming (MILP) decoder \cite{MIALP}, which
requires no information about trapping sets and approaches ML
performance, but with a large decoding complexity.  To deal with the
complexity of the MILP decoder, a multi-stage decoder is proposed in
\cite{Multistage}, where very fast but poor-performing decoders are
combined with the more powerful but much slower MILP decoder.  The
result is a decoder that performs as well as the MILP decoder and with
a high average throughput.  This multi-stage decoder nevertheless
poses considerable practical difficulties for certain applications in
that it requires implementation of multiple decoders, and the
worst-case throughput will be as slow as the MILP decoder.  Our goal
in this paper is to develop decoders that perform much better in the
error floor regime than BP, but with comparable complexity, and no
significant disadvantages.

Our starting point is the iterative ``Divide and Concur'' (DC)
algorithm recently proposed by Gravel and Elser \cite{DAC} for
constraint satisfaction problems. When using DC, one first describes
a problem as a set of variables and local constraints on those
variables.  One then introduces ``replicas'' of the variables; one
replica for each constraint a variable is involved in.\footnote{The
  use of the term ``replica'' in the current context should not be
  confused with the ``replica method'' for averaging over disorder in
  statistical physics, for a review of which we refer the reader
  to~\cite{mezardMontanari09}.}  The DC algorithm then iteratively
performs ``divide'' projections which move the replicas to the values
closest to their current values that also satisfy the local
constraints, and ``concur'' projections which equalize the values of
the different replicas of the same variable. A key idea in the DC
algorithm is to avoid local traps in the dynamics by using the
so-called ``Difference-Map'' (DM) combination of ``divide'' and
``concur'' projections at each iteration.

LDPC codes have a structure that make them a good fit for the DC
algorithm. In fact, Gravel reported on a DC decoder for LDPC codes in
his Ph.D. thesis, although his simulations were very limited in scope 
\cite{GravelThesis}.
We were curious about whether a DC
decoder could be competitive with---or better than---more standard BP
decoders. We were particularly motivated by the idea that the
``traps'' that the DC algorithm's ``Difference-Map'' dynamics
promises to avoid might be related to the ``trapping sets'' that
plague BP decoders of LDPC codes.

To construct a DC decoder, we need to add an important ``energy''
constraint, in addition to the more obvious parity check constraints.
The energy constraint enforces that the correlation between the
channel observations and the desired codeword should be at least some
minimum amount. The effect of this constraint is to ensure that during
the decoding process the candidate solution does not wander too far
from the channel observation.

We found that the DC decoder can be competitive
with BP decoders, but only if many iterations are allowed.
Unfortunately, DC errors are often ``undetected errors'' in that the decoder
returns a codeword that is not the most likely one.  Failures of BP
decoding, in contrast, almost always correspond to failures to converge or
convergence to a non-codeword,
and therefore are detectable.

We show how the DC decoder can be described as a message-passing algorithm.  Using
this formulation, we can see how to import the difference-map idea into
a BP setting. We thus also constructed a novel decoder called the
``difference-map belief propagation'' (DMBP) decoder.  Essentially, DMBP
is a min-sum BP decoder with modified dynamics motivated by the DC
decoder.  Our simulations show that the DMBP decoder improves
performance in the error floor regime quite significantly when
compared with standard sum-product belief propagation (BP) decoders.
We present results for both the additive white Gaussian noise (AWGN)
channel and the binary symmetric channel (BSC).

The rest of the paper is organized as follows.  In Section II, the
DC algorithm is presented, and re-formulated as a message-passing
algorithm.  The DC decoder for LDPC codes is described in Section
III.  The DMBP algorithm is introduced in Section IV.  In Section V we
present simulation results.  Conclusions are given in Section VI.

\section{Divide and concur} \label{DACApp}

In this section, we review Gravel and Elser's ``Divide and Concur''
(DC) algorithm.  Gravel and Elser did {\em not} formulate DC as a message-passing
algorithm, or otherwise compare DC to BP,
but the comparison is illuminating, and helped us design the DMBP
decoder. Thus we present DC in a way that is consistent with Gravel
and Elser's presentation, but makes comparisons to BP easier. We start
by introducing the idea of ``replicas'' in Section~\ref{REPLICASap} in
the context of the familiar alternating projection approach to
constrained satisfaction problems.  In
Section~\ref{DMsec} we introduce and discuss the difference-map
dynamics of DC.  Then, in Section~\ref{DACmp} we reformulate DC as
a message-passing algorithm directly comparable to BP.
  
\subsection{Replicas and alternating projections} \label{REPLICASap}

Consider a system with $N$ variables and $M$ constraints on those
variables.  We seek a configuration of the $N$ variables such that all
$M$ constraints are satisfied. For each constraint that a variable is
involved in, we create one ``replica'' of the variable.  
The idea behind DC is that by constructing a dynamics of
replicas rather than of variables, each constraint can be locally
satisfied (the ``divide'' step), and then later the possibly different
values of replicas of the same variable can be forced to equal each
other (the ``concur'' step).

Denote using $\chkRep{a}$ the vector containing the values of all the
replicas associated with the $a$th constraint and let $\varRep{i}$ be the
vector of all the values of replicas associated with the $i$th
variable. Let $\replicas$ be the vector containing all the values of
replicas of all the variables. Now $\chkRep{a}$ for $a=1, 2, \cdots,
M$ and $\varRep{i}$ for $i=1,2,\cdots, N$ are two different ways to
partition $\replicas$ into mutually exclusive sets. 

There are two projection operations, the ``divide'' projection and the
``concur'' projection, denoted by $P_D$ and $P_C$, respectively.  Both
projections act on $\replicas$ and output a new $\replicas$ that
satisfies certain requirements. Since $\replicas$ can be partitioned
into mutually exclusive sets, the projections are actually applied to
each set independently.  The divide projection is a product
of local divide projections $P_D^a(\chkRep{a})$ that operate on each
$\chkRep{a}$ for $a=1, 2, \cdots, M$. If $\chkRep{a}$ satisfies the
$a$th constraint, $P_D^a(\chkRep{a}) = \chkRep{a}$; otherwise,
$P_D^a(\chkRep{a}) = \tilde{\textit{\textbf{r}}}_{(a)}$ such that
$\chkRepTil{a}$ is the closest vector to $\chkRep{a}$ that satisfies
the $a$th constraint. The metric used is normally ordinary
Euclidean distance.

The divide projection forces all constraints to be satisfied, but has
the effect that replicas of the same variable do not necessarily agree
with one another. The concur projection is a product of local
concur projections $P_C^i(\varRep{i})$ that act on $\varRep{i}$ for
$i=1,2,\cdots, N$. Let $\varAvg{i}$ be the average of all the elements
in $\varRep{i}$ and construct a vector $\varRepAvg{i}$ with each
element equal to $\varAvg{i}$, with dimensionality the same as $\varRep{i}$. Then
$P_C^i(\varRep{i})=\varRepAvg{i}$. While the concur projection equalizes
the values of the replicas of the same variable, the new values of the
replicas may violate some constraints.

The overall projection $P_D(\replicas)$ [alternately $P_C(\replicas)$]
is defined as applying $P_D^a(\cdot)$ [$P_C^i(\cdot)$] to $\chkRep{a}$ for
$a = 1, 2, \ldots, M$ [$\varRep{i}$~for $i = 1, 2, \ldots, N$].  The
$M$ [$N$] output vectors are then reassembled into the updated
$\replicas$ vector through appropriate ordering.

A strategy is needed to combine these two projections to find a set of
replica values such that all constraints are satisfied and all
replicas of the same variable are equal.  The simplest approach is to
alternate two projections, i.e., $\replicas_{t+1} =
P_C(P_D(\replicas_t))$, where $\replicas_{t}$ is the vector of
  replica values at the $t$th iteration. This scheme works well for
  convex constraints, but it is prone to getting stuck in short cycles
  (``traps'') that do not correspond to solutions.

\begin{figure}
  \centering
  \includegraphics[width=0.35\textwidth]{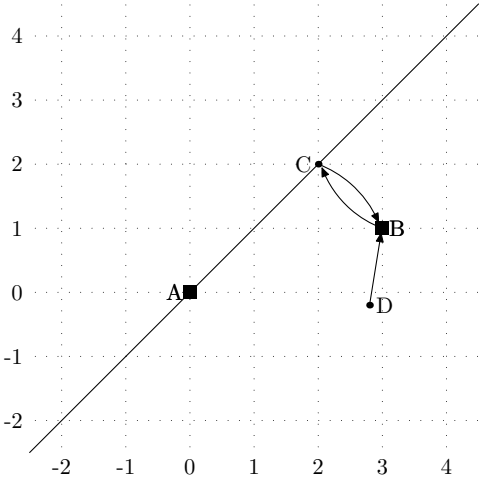}
  \caption {A simple example of a trap in an iterated projection
    strategy. If one iteratively projects to the nearest point that
    satisfies the constraints ($A$ or $B$), and then the nearest point
    where the replica values are equal (the diagonal line) one may be
    trapped in a short cycle ($B$ to $C$ to $B$ and so on) and never
    find the true solution at point $A$.}
\label{trap}
\end{figure}

To illustrate this point, consider the situation shown in
Fig. \ref{trap}, where we imagine that the space of replicas of a particular
variable is only
two-dimensional, i.e., the variable in question participates in two constraints. 
The diagonal line represents the
requirement that all replicas are equal, since they are replicas of the
same variable. The points $A$ and $B$
are the two pairs of replica values that satisfy the variable's constraints. The
only common value that the replicas can take that satisfies both constraints is
zero, i.e. point $A$. 
However, if one initializes replica values near point
$B$, say at $D$, and applies the divide projection, 
then one will move to $B$, the nearest point that
satisfies the constraints. Next, the concur projection will move
to point C, the nearest point (along the diagonal) where the replica values
are equal. Continued application of divide and concur projections, in sequence,
moves the system to $B$, then back to $C$, then back to $B$, and so forth.
Alternating projections cause the system to be stuck in a simple trap.
Of course, this is only a toy two-dimensional example, but in
non-convex high-dimensional spaces it is plausible that an iterated
projection strategy is prone to falling into such traps.

\subsection{Difference Map} \label{DMsec}

The difference map (DM) is a strategy that improves alternating
projections by turning traps in the dynamics into repellers. It is
defined by Gravel and Elser as follows:
\begin{eqnarray}\label{DM}
  \replicas_{t+1} = \replicas_t +
  \beta\left[P_C(f_D(\replicas_{t}))-P_D(f_C(\replicas_{t}))\right]
\end{eqnarray}
where $f_s(\replicas_{t}) = (1+\gamma_s)P_s(\replicas_{t})-\gamma_s
\replicas_{t}$ for $s=C$ or $D$ with $\gamma_C=-1/\beta$ and
$\gamma_D=1/\beta$. The parameter $\beta$ can be chosen to optimize
performance. 

We focus here exclusively on the case $\beta=1$, which is usually an
excellent choice and corresponds to what Fienup called the ``hybrid
input-output'' algorithm, originally developed in the
context of image reconstruction \cite{Fienup}\cite{ElserPNAS}. See \cite{Bauschke}
for a review of Fienup's algorithm and other projection algorithms for image
reconstruction, and their relationship with earlier convex
optimization methods.

For $\beta = 1$, the dynamics~(\ref{DM}) simplify to
\begin{eqnarray}\label{DMbeta2}
\replicas_{t+1} = P_C\big(\replicas_t +
2[P_D(\replicas_{t})-
\replicas_{t}]\big)-[P_D(\replicas_{t})-\replicas_t].
\end{eqnarray}
It can be proved that if a fixed point in the dynamics
$\replicas^{*}$ is reached, i.e., $\replicas_{t+1} =
\replicas_t=\replicas^{*}$, then that fixed point must {\em correspond} to a
solution of the problem.
It is important to note that the fixed point itself is {\em not necessarily}
a solution.
The solution $\replicas_{sol}$ corresponding to a fixed point $\replicas^{*}$
can be obtained using 
$\replicas_{sol} = P_D(\replicas^{*})$ or $\replicas_{sol} =
P_C(\replicas^{*} + 2[P_D(\replicas^*)-\replicas^*])$.

We have found it very useful to think of the difference-map dynamics
for a single iteration as breaking down into a three-step process.  The expression
$[P_D(\replicas_{t})- \replicas_{t}]$ represents the change to the
current values of the replicas resulting from the divide projection.
In the first step, the values of the replicas move {\em twice} the
desired amount indicated by the divide projection. We refer to these
new values of the replicas as the ``overshoot'' values
$\textit{\textbf{r}\/}_t^{over} = \replicas_t + 2[P_D(\replicas_{t})
-\replicas_{t}]$. Next the concur projection is applied to the
overshoot values to obtain the ``concurred'' values of the replicas
$\textit{\textbf{r}\/}_t^{conc} =
P_C(\textit{\textbf{r}\/}_t^{over})$.  Finally the overshoot, i.e.,
the extra motion in the first step, is subtracted from the concur
projection result to obtain the replica value for the next iteration
$\replicas_{t+1}=
\textit{\textbf{r}\/}_t^{conc}-[P_D(\replicas_{t})-\replicas_t]$.

In Fig.~\ref{trap2} we return to our previous example and see that the
DM dynamics do not get stuck in a trap.  Suppose, as before, that
point $A$ is at $(0,0)$, point $B$ is at $(3,1)$, and and that we now start
initially at point $\replicas_{1}= (2,2)$.  The divide projection
would take us to point $B$, but the overshoot takes us twice as far to
$\replicas_1^{over}= (4,0)$. The concur projection takes us back to
$\replicas_1^{conc}=(2,2)$.  Finally, the overshoot is corrected so
that $\replicas_2=(1,3)$.  The next full iteration takes us to
$\replicas_3=(0,4)$ (sub-steps are tabulated in Fig.~\ref{trap2}).
Now however, we are closer to $A$ then to $B$.  Therefore, the next
overshoot take us to $\replicas_3^{over}= (0,-4)$, from which we would
move to $\replicas_3^{conc}=(-2,-2)$, and $\replicas_4
=\replicas^*=(-2,2)$.  Finally, at $\replicas_4$ we have reached a fixed point in
the dynamics that corresponds to the solution at $A$ (which can be obtained from
the final value of $P_D(\replicas_t)$ or $\replicas_t^{conc}$).

\begin{figure}
\centering
\begin{minipage}[c]{2.75in}
  \includegraphics[width=3in]{figures/dmbp.2}
  \end{minipage}
\vspace{0.2in}
\begin{minipage}[c]{2.75in}
\vspace{0.3in}
\begin{tabular}{c|c|c|c|c}
$t$ & $\replicas_t$ & $P_D(\replicas_t)$ & 
$\replicas_t^{over}$ & 
$\replicas_t^{conc}$ \\ \hline
$1$ & $(2,2)$ & $(3,1)$& $(4,0)$ & $(2,2)$ \\ \hline 
$2$ & $(1,3)$ & $(3,1)$ & $(5, -1)$ & $(2,2)$ \\ \hline
$3$ & $(0,4)$ & $(0,0)$ & $(0, -4)$ & $(-2, -2)$ \\ \hline
$4$ & $(-2,2)$& $(0,0)$ & $(2,-2)$ & $(0,0)$ \\ \hline
$5$ & $(-2,2)$&&&
\end{tabular}
\end{minipage}
\caption {An example showing how DM dynamics avoids traps. If we start
  at the point $r_1$, an iterated projections dynamics would be
  trapped between point $B$ and $r_1$, and never find the solution at
  $A$. DM dynamics will instead be repelled from the trap and move to
  $r_2$ (via the three sub-steps denoted with dashed lines 
$r_1^{over}$, $r_1^{conc}= r_1$, and $r_2$), 
then move to $r_3$, and then end at the fixed point $r_4=r^*$,
  which corresponds to the solution at $A$.}
\label{trap2}
\end{figure}

We can generalize from this example to understand how the DM dynamics
turns a trap into a ``repeller,'' where at each iteration, one moves
away from the repeller by an amount equal to the distance between the
constraint involved and the nearest point that satisfies the
requirement that the replicas be equal. Of course, DM dynamics are not
a panacea; it is possible that DC can get caught in more complicated
cycles or ``strange attractors'' and never find an existing solution;
but least it will does not get caught in simple traps.

\subsection{DC as a message-passing algorithm}\label{DACmp}

We now turn to developing an alternative interpretation of DC, as a
message-passing algorithm on a graph.  ``Messages'' and ``beliefs''
are similar to those in BP, but message-update and belief-update rules
are different.  To begin with, we construct a bi-partite ``constraint
graph'' of variable nodes and constraint nodes, where each variable is
connected to the constraints it is involved in. A constraint graph can
be thought of as a special case of a factor graph \cite{FactorGraph},
where each allowed configuration is given the same weight, and
disallowed configurations are given zero weight.

We identify the DC ``replicas'' with the edges of the graph. We
denote by $\varChkRep{i}{a}(t)$ the value of the replica on the edge
joining variable $i$ to constraint $a$ at the beginning of iteration
$t$, i.e., the appropriate element of $\varRep{i}(t)$.  We similarly
denote by $\varChkRepOver{i}{a}(t)$ and $\varChkRepConc{i}{a}(t)$ the
``overshoot'' and ``concurred'' values of the same replica.  We note
that these are all scalars.

We can alternatively think of the initial value of a replica
$\varChkRep{i}{a}(t)$ as a ``message'' from the variable node $i$ to
the constraint node $a$ that we denote as $m_{i \rightarrow a}(t)$.
The set of incoming messages to constraint node $a$,
$\textit{\textbf{m}}_{\rightarrow a}(t) \equiv \{m_{i \rightarrow
  a}(t) : i \in \mbox{$\cal N$}(a)\}$ where $\mbox{$\cal N$}(a)$ is
the set of variable indexes involved in constraint $a$, can therefore
be expressed as $\textit{\textbf{m}}_{\rightarrow a}(t) =
\chkRep{a}(t)$.

In the three-step interpretation of the DM dynamics described above,
these replica values are next transformed into overshoot values by
moving by twice the amount indicated by the divide projection. Because
the overshoot values are computed locally at a constraint node using
the messages into to the constraint node, we can think of the overshoot
values $\varChkRepOver{i}{a}(t)$ as messages from the constraint node
$a$ to their neighboring variable nodes $i$, denoted by $m_{a
  \rightarrow i}(t)$.  The set of outgoing messages from constraint
node $a$ is $\textit{\textbf{m}}_{a \rightarrow}(t) \equiv \{m_{a
  \rightarrow i}(t) : i \in \mbox{$\cal N$}(a)\}$.  This set can thus
be calculated as $\textit{\textbf{m}}_{a \rightarrow}(t) =
\replicas_{a}^{over}(t) = \chkRep{a}(t) + 2[P_D^a(\chkRep{a}(t)) -
\chkRep{a}(t)] = \textit{\textbf{m}}_{\rightarrow a}(t) + 2[ P_D^a(
\textit{\textbf{m}}_{\rightarrow a}(t))-
\textit{\textbf{m}}_{\rightarrow a}(t)] $.

The next step of the DC algorithm takes the overshoot replica values
$\varChkRepOver{i}{a}(t)$ and computes concurred values
$\varChkRepConc{i}{a}(t)$ using the concur projection.  Note that the
concurred values for replicas that are connected to the same variable
node $i$ are all equal to each other.  We can think of these concurred
values as ``beliefs,'' denoted by $b_i(t)$. Just as in BP, the beliefs
at a variable node $i$ are computed using all the messages coming into
that variable node. However, while the BP belief is a sum of incoming messages,
the DC belief is an average:
\begin{equation}
b_i(t) = P_C^i(\replicas_{[i]}^{\over} (t)) =
\frac{1}{|\mbox{$\cal M$}(i)|} \sum_{a \in \mbox{$\cal M$} (i)} m_{a \rightarrow i}(t) 
\end{equation}
where $\mbox{$\cal M$}(i)$ is the set of constraint
indexes in which variable $i$ participates.

Finally, the DC rule for computing the new replica values at the
next iteration is to take the concurred values and subtract a
correction for the amount we overshot when we computed the overshot
values. In terms of our belief and message formulation, we compute the
outgoing messages from a variable node at the next iteration using the
rule
\begin{equation}\label{overshootcorrect}
  m_{i \rightarrow a}(t+1) = b_i(t) - 
  \frac{1}{2} \left[m_{a \rightarrow i}(t) - m_{i \rightarrow a}(t)\right].
\end{equation}
Comparing with the ordinary BP rule
\begin{equation}
  m_{i \rightarrow a}(t+1) = b_i(t) - m_{a \rightarrow i}(t),
\end{equation}
we note that the message out of a variable node in DC also depends
on the value of the same message at the previous iteration, which is not
the case in BP.

To summarize, the overall structure of BP and DC as message-passing
algorithms is similar.  In both one iteratively updates beliefs at
variable nodes and messages between variable nodes and constraint
nodes.  Furthermore, messages out of a constraint node are computed
based on the messages into the constraint node, beliefs are computed
based on the messages into a variable node, and the messages out of
the variable node depend on the beliefs and the messages into a
variable node. The differences are in the specific forms of the
message-update and belief-update rules, and the fact that a
message-update rule for a message out of a variable node in DC also
depends on the value of the same message in the previous iteration.

\section{DC decoder for LDPC codes} \label{DCdecDefs}

Decoding of LDPC codes can be described as a constraint satisfaction
problem. We restrict ourselves here to binary LDPC codes, although
generalizations to $q$-ary codes are straightforward.  Searching for a
codeword is equivalent to seeking a binary sequence which satisfies
all the single-parity check (SPC) constraints simultaneously. We also
add one important additional constraint, which is that the likelihood
of a binary sequence must be greater than some minimum amount.  Then
the decoding problem can be divided into many simple sub-problems which
can be solved independently using the DC approach.

Let $M$ and $N$ be the number of SPC constraints and bits of a binary
LDPC code, respectively. Let $\textit{\textbf{H}}$ be the parity check
matrix which defines the code. Assume BPSK signaling with unit energy,
which maps a binary codeword
$\textit{\textbf{c}}=(c_{1},c_{2},\ldots,c_{N})$ into a sequence
$\textit{\textbf{x}}=(x_{1},x_{2},\ldots,x_{N})$, according to
$x_{i}=1-2c_{i}$, for $i=1,2,\ldots,N$. The sequence
$\textit{\textbf{x}}$ is transmitted through a channel and the
received channel observations are denoted
$\textit{\textbf{y}}=(y_{1},y_{2},\ldots,y_{N})$.  Let the
log-likelihood ratios (LLR's) corresponding to the received channel
observations be $\textit{\textbf{L}}=(L_{1},L_{2},\ldots,L_{N})$,
where
\begin{eqnarray}
  L_i = \log \left( \frac{\Pr[y_i | x_i = 1]}{\Pr[y_i | x_i = -1]} \right).
  \nonumber
\end{eqnarray}
Our goal is to recover the transmitted sequence of variables
$\textit{\textbf{x}}$. To do this, we will search for a sequence of
$\pm 1$'s that satisfies all the SPC constraints and has the highest
likelihood or, equivalently, the lowest ``energy,'' where the energy
is defined as $E = - \sum_{i=1}^N L_i x_i.$ Note that although our
desired sequence consists only of $\pm 1$ variables, the ``replica''
values, or equivalently ``messages'' and ``beliefs,'' are real-valued.

In all, we have $N$ variables $x_k$, and $M+1$ constraints, of which
$M$ are SPC constraints, with one additional energy constraint.  We
will write the energy constraint as $-\sum_{i} L_i x_i \le E_{\max}$,
where different choices of $E_{\max}$ result in different decoders. It
is not obvious how to choose $E_{\max}$; we performed preliminary
experiments to search for an $E_{\max}$ that optimizes decoding
performance. Somewhat surprisingly, the best choice for $E_{\max}$ is
one that for which the energy constraint can never actually be
satisfied: we found that $E_{\max} = -(1+\epsilon) \sum_i |L_i|$, with
$0 < \epsilon \ll 1$ was an excellent
choice.  The fact that the energy constraint is never satisfied is not
a problem because the decoder terminates if it finds a codeword that
satisfies all the SPC constraints. Until then, the effect of the energy constraint
is to keep the replica values near the transmitted
sequence.

We will describe the DC decoder as an iterative message-update
algorithm on a constraint graph, following the formulation in section
\ref{DACmp}. We use $N$ variable indexes $i=1,2,\cdots,N$ and $M+1$
constraint indexes $a=0,1,2,\cdots,M$, where the $0$th constraint is
the energy constraint. SPC
constraints involve a small number of variables, but the energy
constraint involves every variable.  
To lay the groundwork for the overall DC decoder, we now
explain how to perform the divide and concur projections.

\subsection{Divide and concur projections for LDPC decoding}

The divide projection $P_D$ can be partitioned into a collection of
$M+1$ projections $P_D^a$, where each projection operates independently on a vector
of messages $\textit{\textbf{m}}_{\rightarrow a}(t) \equiv \{m_{i
  \rightarrow a}(t) : i \in \mbox{$\cal N$}(a)\}$ and outputs a vector
(of the same dimensionality) of projected messages
$P_D^a(\textit{\textbf{m}}_{\rightarrow a}(t))$.  The output vector is
as close as possible to the original values
$\textit{\textbf{m}}_{\rightarrow a}(t)$ while satisfying the $a$th
constraint.  

The SPC constraints require that the variables involved in a
constraint are all $\pm 1$, with an even number of $-1$'s.  For these
constraints we efficiently perform the divide projection as follows:
\begin{itemize}
\item Make a hard decision $h_{ia}$ on each of $m_{i \rightarrow
    a}(t)$ such that $h_{ia} = 1$ if $m_{i \rightarrow a}(t) > 0$,
  $h_{ia} = -1$ if $m_{i \rightarrow a}(t) < 0$, and $h_{ia}$ is chosen
  to be $1$ or $-1$ randomly if $m_{i \rightarrow a}(t) = 0$.
\item Check if $\textit{\textbf{h}}_a$ contains an even number of
  $-1$'s. If it does, set $P_D^a(\textit{\textbf{m}}_{\rightarrow
    a}(t))=\textit{\textbf{h}}_a$ and return.
\item Otherwise, let $\nu = \mathop{\textmd{argmin}}_{i} |m_{i
    \rightarrow a}(t)|$.  Especially for the BSC, it is possible that
  several messages have equally minimal $|m_{i \rightarrow a}(t)|$. In
  this case, we randomly pick one of them and use its index as $\nu$.
\item Flip $h_{\nu a}$, i.e., if $h_{\nu a}=-1$, set it to $1$ and if
  $h_{\nu a}=1$, set it to $-1$. Then set
  $P_D^a(\textit{\textbf{m}}_{\rightarrow a}(t))=\textit{\textbf{h}}_a$
  and return.
\end{itemize} 

Recall that the energy constraint is $- \sum_{i=1}^{N} x_i L_i \le
E_{\max}$. This implies a divide projection on the vector of messages
$\textit{\textbf{m}}_{\rightarrow 0}(t)$, performed as follows:
\begin{itemize}
\item If the energy constraint is already satisfied by the messages
  $\textit{\textbf{m}}_{\rightarrow 0}(t)$, return the current
  messages, i.e., $P_D^0(\textit{\textbf{m}}_{\rightarrow
    0}(t))=\textit{\textbf{m}}_{\rightarrow 0}(t)$.  (Recall however
  that the energy constraint will never be satisfied for the choice of
  $E_{\max} = - (1 + \epsilon) \sum_i |L_i|$ that we use in our simulations.)
\item Otherwise, find $\textit{\textbf{h}}_0$ which is the closest
  vector to $\textit{\textbf{m}}_{\rightarrow 0}(t)$ and satisfies the
  energy constraint. An easy application of vector calculus can be
  used to derive that the $i$th component $h_{i0}$ is given by the
  formula
\begin{equation}
  h_{i0} = m_{i \rightarrow 0}(t) - 
  \frac{L_i (\sum_i L_i m_{i \rightarrow 0}(t)+E_{\max})}{\sum_i L_i^2}
\end{equation}
Set $P_D^0(\textit{\textbf{m}}_{\rightarrow
  0}(t))=\textit{\textbf{h}}_0$ and return.
\end{itemize}

Finally, the concur projection $P_C$ can be partitioned into a set of
$N$ projection operators $P_C^i$, 
where each $P_C^i$ operates independently on the vector
of messages $\textit{\textbf{m}}_{\rightarrow i} \equiv \{m_{a
  \rightarrow i}(t) : a \in \mbox{$\cal M$}(i)\}$ and outputs the
belief $b_i(t)$, the average over the components of the vector
$\textit{\textbf{m}}_{\rightarrow i}$.

\subsection{DC algorithm for LDPC decoding}

The overall DC decoder proceeds as follows.

\begin{itemize}
\item[] \hspace{-1.5em}{\bf 0. Initialization:} Set the maximum number
  of iterations to $T_{\max}$ and the current iteration to $t=1$.
  Initialize the messages out of variable nodes $m_{i \rightarrow
    a}(t=1)$ for all $i$ and $a \in \mbox{$\cal M$}(i)$ to equal
  $2p_i-1$, where $p_i$ is the {\it a priori} probability that the
  $i$th transmitted symbol $x_i$ was a $1$, given by $p_i \equiv
  \exp(L_i)/(1+\exp(L_i))$.
\item[] \hspace{-1.5em}{\bf 1. Update messages from checks to
    variables:} Given the messages $\textit{\textbf{m}}_{\rightarrow
    a}(t) \equiv \{m_{i \rightarrow a}(t) : i \in \mbox{$\cal
    N$}(a)\}$ into each constraint $a$, compute the messages out of
  each constraint $\textit{\textbf{m}}_{a \rightarrow}(t) \equiv
  \{m_{a \rightarrow i}(t) : i \in \mbox{$\cal N$}(a)\}$ using the
  overshoot formula
\begin{equation}
  \textit{\textbf{m}}_{a \rightarrow}(t) = \textit{\textbf{m}}_{\rightarrow a}(t) + 
  2[P_D^a(\textit{\textbf{m}}_{\rightarrow a}(t)) - 
  \textit{\textbf{m}}_{\rightarrow a}(t)]
\end{equation}
where $P_D^a(\textit{\textbf{m}}_{\rightarrow a}(t))$ is the divide
projection operation for constraint $a$.
\item[] \hspace{-1.5em}{\bf 2. Update beliefs:} Compute the beliefs at
  each variable node $i$ using the concur projections
\begin{equation}
  b_i(t) = P_C^i(\textit{\textbf{m}}_{\rightarrow i}(t)) = 
  \frac{1}{|\mbox{$\cal M$}(i)|} \sum_{a \in \mbox{$\cal M$}(i)} m_{a \rightarrow i}(t).
\end{equation}
\item[] \hspace{-1.5em}{\bf 3. Check if codeword has been found:}
  Create $\hat{\textit{\textbf{c}}}=\{\hat{c}_i\}$ such that
  $\hat{c}_i = 1$ if $b_i(t) <0$, $\hat{c}_i =0$ if $b_i(t) > 0$ and
  flip a coin to decide $\hat{c}_i$ if $b_i(t) =0$. If
  $\textit{\textbf{H}}\hat{\textit{\textbf{c}}}=\textbf{0}$ output
  $\hat{\textit{\textbf{c}}}$ as the decoded codeword and stop.
\item[] \hspace{-1.5em}{\bf 4. Update messages from variables to
    checks:} Increment $t := t+1$. If $t > T_{\max}$ stop and return
  {\tt FAILURE}. Otherwise, update each message out of the variable
  nodes using the ``overshoot correction'' rule given in equation
  (\ref{overshootcorrect}) and go back to Step 1.
\end{itemize}

As already mentioned in the introduction, the DC decoder performs
reasonably well, but with some problems. We defer a detailed
discussion of the DC simulation results until section
\ref{simulations}.  First we describe a second and novel decoder, the
difference-map belief propagation (DMBP) decoder.

\section{DMBP Decoder}

Our motivation in creating the DMBP decoder was that BP decoders
generally perform well, but they seem to use something like an
iterated projection strategy, and perhaps the trapping sets that
plague the error-floor regime are related to the ``traps'' that the
difference-map dynamics is supposed to ameliorate. Since we can also
describe DC decoders as message-passing decoders, we could try to
create a new BP decoder that was a mixture of BP and difference-map
ideas.

For simplicity, we work with a min-sum BP decoder using messages and
beliefs that correspond to log-likelihood ratios. Note that the
min-sum message update rule is much simpler to implement in hardware
than the standard sum-product rule. Normally, sum-product (or some
approximation to sum-product) BP decoders are favored over min-sum BP
decoders because they perform better, but we found that the
straightforward min-sum DMBP decoder will out-perform the more
complicated sum-product BP decoder.  Our preliminary simulations also
show, somewhat surprisingly, that the min-sum DMBP decoder slightly
out-performs a sum-product DMBP decoder.  (We don't further discuss
the sum-product DMBP decoder herein.)

We use the same notation for messages and beliefs that were used in
the discussion of the DC decoder in Section~\ref{DCdecDefs}.  We
compare, on an intuitive level, the min-sum BP decoder with the DC
decoder in terms of belief updates and message-updates at both the
variable and check nodes.

Beginning with the message-updates at a check node, the standard
min-sum BP update rules are to take incoming messages $m_{i
  \rightarrow a}(t)$ and compute outgoing messages according to the
rule that
\begin{equation}\label{minsumupdate}
  m_{a \rightarrow i}(t) = \left(\min_{j \in \,\mbox{$\cal N$}(a) \backslash i}
    |m_{j \rightarrow a}(t)|\right)
  \prod_{j \in \,\mbox{$\cal N$}(a) \backslash i}\textmd{sgn}(m_{j \rightarrow a}(t)),
\end{equation}
where $\textmd{sgn}(z) = z/|z|$ if $z \ne 0$, and $\textmd{sgn}(z) = 0$ if 
$z = 0$.  Comparing
with the DC ``overshoot'' message-update rule, we note that the
min-sum updates, in some sense, also ``overshoot''.  For example, at a
check node that has three incoming positive messages and one incoming
negative message, we obtain three outgoing negative messages and one
outgoing positive message.  This overshoots the ``correct'' solution
of having an even number of negative messages (since the parity
check must ultimately be connected to an even number of variables with value
$-1$). Because the min-sum
rule for messages outgoing towards a particular variable ignore the
incoming message from that variable, all the outgoing messages move
beyond what is necessary (at least in terms of sign) to satisfy the
constraint.  Since we {\em want} an overshoot, we decided to leave this rule
unmodified.

Turning to the belief update rule, the standard BP rule is to compute
the belief as the {\em sum} of incoming messages (including the
message from the observation), while the DC rule is that the belief
is the {\em average} of incoming messages. We decided to use the
compromise rule
\begin{equation}\label{beliefupdate}
  b_i(t) = Z \left( L_i + \sum_{a \in \mbox{$\cal M$}(i)}m_{a \rightarrow i}(t) \right)
\end{equation}
where $Z$ is a parameter chosen by optimizing decoder performance.

Finally, for the message-update rule for messages at the variable
nodes, we directly copy the ``correction'' rule from DC. Our
intuitive idea is that perhaps standard BP is missing the correction
that is important in repelling DM dynamics from traps.

To summarize, the DMBP decoder works as follows:

\begin{itemize}
\item[] \hspace{-1.8em} {\bf 0. Initialization:} Set the maximum
  number of iterations to $T_{\max}$ and the current iteration to
  $t=1$.  Initialize the the messages out of variable nodes $m_{i
    \rightarrow a}(t=1)$ for all $i$ and $a \in \mbox{$\cal M$}(i)$ to
  equal $L_i$.
\item[] \hspace{-1.5em}{\bf 1. Update messages from checks to
    variables:} Given the messages $m_{i \rightarrow a}(t)$ coming
  into the constraint node $a$, compute the outgoing messages using
  the min-sum message update rule given in equation
  (\ref{minsumupdate}).
\item[] \hspace{-1.5em}{\bf 2. Update beliefs:} Compute the beliefs at
  each variable node $i$ using the belief update rule given in
  equation (\ref{beliefupdate}).
\item[] \hspace{-1.5em}{\bf 3. Check if codeword has been found:}
  Create $\hat{\textit{\textbf{c}}}=\{\hat{c}_i\}$ such that
  $\hat{c}_i = 1$ if $b_i(t) <0$, $\hat{c}_i =0$ if $b_i(t) > 0$ and
  flip a coin to decide $\hat{c}_i$ if $b_i(t) =0$. If
  $\textit{\textbf{H}}\hat{\textit{\textbf{c}}}=\textbf{0}$ output
  $\hat{\textit{\textbf{c}}}$ as the decoded codeword and stop.
\item[] \hspace{-1.5em}{\bf 4. Update messages from variables to
    checks:} Increment $t := t+1$. If $t > T_{\max}$ stop and return
  {\tt FAILURE}. Otherwise, update each message out of the variable
  nodes using the ``overshoot correction'' rule given in equation
  (\ref{overshootcorrect}) and go back to Step 1.
\end{itemize}

\section{Simulation results}\label{simulations}

In this section, we compare simulation results of the DC and DMBP
decoders to those of a variety of other decoders. The decoding
algorithms are applied to two kinds of LDPC codes and simulated over
both the BSC and the AWGN channel. One code is a random regular LDPC
code with length 1057 and rate 0.77, obtained from
\cite{DataBase}. The other code is a quasi-cyclic (QC) ``array'' LDPC code
\cite{Fan}\cite{Dolecek} with length
2209 and rate 0.916.

The first point of comparison of our proposed decoders is to
sum-product BP decoding.  When simulating transmission over the BSC,
in order better to probe the error floor region, we implement the
multistage decoder introduced in~\cite{Multistage}.  Multistage
decoders pre-append simpler decoders (in our case Richardson \&
Urbanke's Algorithm-E~\cite{richardsonUrbanke:01} and/or regular
sum-product BP) to the more complex decoders of interest (e.g., DC).
The simpler decoders either decode or fail to decode in a detectable
way (e.g., by not converging in BP's case).  Failures to decode
trigger the use of the more complex decoders.  In this way one can
often achieve the WER performance of the most complex decoder at an
expected complexity close to that of the most simple decoder.  Our
first use of the multistage approach in this paper is to calculate the
performance of sum-product BP decoding for the BSC.  We implement a
multistage decoder that combines a first-stage Algorithm-E to a
second-stage sum-product BP.  We term the combination E-BP.  For the
sum-product BP simulations of the AWGN channel simulations we
implement a standard sum-product BP decoder (and not a multistage
decoder) as we have found Algorithm-E has very poor performance on the
AWGN channel and thus does not appreciably reduce simulation time.

For DC and DMBP we provide results for standard (single-stage)
implementations of both algorithms as well as for multi-stage
implementations.  As per the discussion above, we use E-BP as the
initial stages for simulations over the BSC and BP by itself as a
first stage for simulations of the AWGN channel.  We denote the
resulting multi-stage decoders by E-BP-DMBP, E-BP-DC, BP-DMBP and
BP-DC.

Our final points of comparison are to linear programming (LP) decoding
and mixed-integer LP (MILP) decoding.  
Our LP decoders were accelerated using
Taghavi and Siegel's ``adaptive'' methods~\cite{Taghavi}, 
and ultimately relied on the simplex
algorithm as implemented in the GLPK linear programming library \cite{GLPK}. 
For the BSC, we implement the
multistage decoders E-BP-LP and E-BP-MILP($l$) for $l=10$, where $l$
is the maximum number of integer (in fact binary) constraints the MILP
decoder is allowed. 
Further details of these decoders and results can
be found in~\cite{Multistage}.

Regarding the decoding parameters of our new algorithms, for the
random LDPC code, we use $Z=0.35$ for the DMBP decoder over both BSC
and the AWGN channel.  For the array code, we use $Z=0.405$ over the
BSC and $Z=0.445$ over the AWGN channel.  

Finally, we are often able to estimate a lower bound on the word
error rate (WER) of ML decoding.  When our decoders return a codeword
that is different from the transmitted codeword, but has a higher
probability, we know that an optimal ML decoder would also have made a
decoding ``error.''  The proportion of such events provides an estimated lower bound
on ML performance. (The true ML WER could be above the lower bound 
because an ML decoder
may also make errors on blocks for which our decoder fails to converge, events
that our estimate assumes ML would decode correctly.)

\begin{figure}[t]
\begin{center}
\subfigure[Results when $T_{\max} = 50$
    iterations]{\includegraphics[width=3.5in]{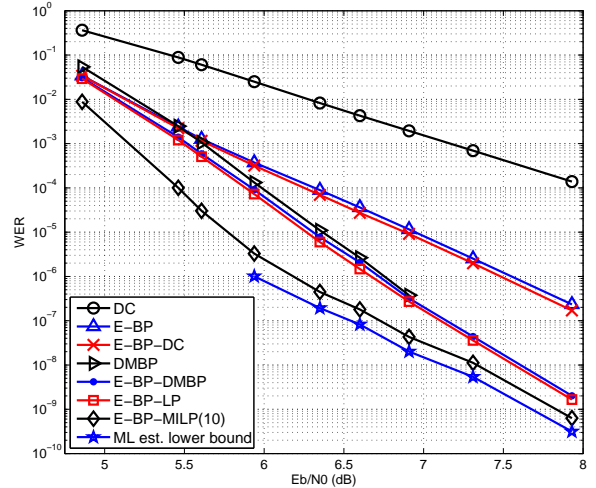} 
      \label{N1057BSC_a}}
\quad
\subfigure[Results when $T_{\max} = 300$
    iterations]{\includegraphics[width=3.5in]{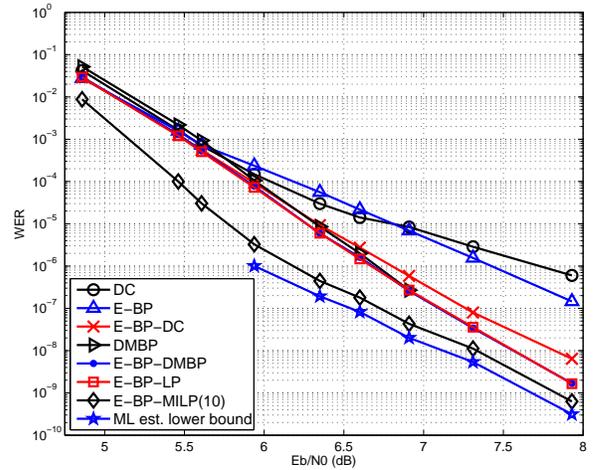}
      \label{N1057BSC_b}}
\end{center}
  \caption {Error performance comparisons for a length-1057, rate-0.77
    random LDPC code over the BSC.} \label{N1057BSC}
\end{figure}

Figure \ref{N1057BSC} plots the word error rates of the various
algorithms for the length-1057 random LDPC code when transmitted over
the BSC.  We plot WER versus SNR, assuming that the BSC results from
hard-decision demodulation of a BPSK $\pm 1$ sequence
transmitted over an AWGN channel.  The resulting relation between the
crossover probability $p$ of the equivalent BSC-$p$ and the SNR of the
AWGN channel is
$
p = \mbox{Q}\left(\sqrt{2 R \cdot 10^{SNR/10}}\right), \nonumber
$
where $R$ is the rate of the code and $\mbox{Q}(\cdot)$ is the Q-function.  In
Figure~\ref{N1057BSC_a} we plot results when all iterative algorithms
are limited to $T_{\max}=50$ iterations, and in
Figure~\ref{N1057BSC_b} to $T_{\max}=300$ iterations.  We observe that
E-BP-DMBP improves the error floor performance dramatically compared
with E-BP (E-BP-DC also improves significantly compared with E-BP if
one allows for 300 iterations) and in the high SNR region E-BP-DMBP
with 50 iterations is very close to the estimated lower bound of the
maximum likelihood (ML) decoder.  Note also that a pure DMBP decoder
has almost the same performance as E-BP-DMBP for both 50 and 300
iterations, so the E-BP-DMBP performance in the very high SNR regime
should be indicative of the pure DMBP performance.


From Figure \ref{N1057BSC}, we also observe that the pure DC decoder
needs many more iterations to obtain good performance compared with
both BP and DMBP. For 300 iterations, DC performs better than E-BP
at lower SNR, but exhibits an apparent error floor as the SNR
increases.  This high error floor is mostly the result of the DC
decoder returning a codeword with {\em lower} probability than the
transmitted codeword.  For example, for an SNR of 6.60 dB, 80\% of
DC errors are of this type, while for an SNR of 7.31 dB, the
percentage rises to 98\%. In contrast, the BP and DMBP decoders
essentially never make this kind of error.

Notice that E-BP-LP has a very similar performance to DMBP, and also
that E-BP-MILP with 10 fixed bits performs the best among all the
decoders and almost approaches the estimated ML lower bound. However,
DMBP decoders should be significantly more practical to construct in
hardware, because they are message-passing decoders similar to
existing BP decoders, while LP and MILP decoders do not currently have
efficient and hardware-friendly message-passing implementations.

\begin{figure}[t]
\begin{center}
\subfigure[Results when $T_{\max} = 50$
    iterations]{\includegraphics[width=3.5in]{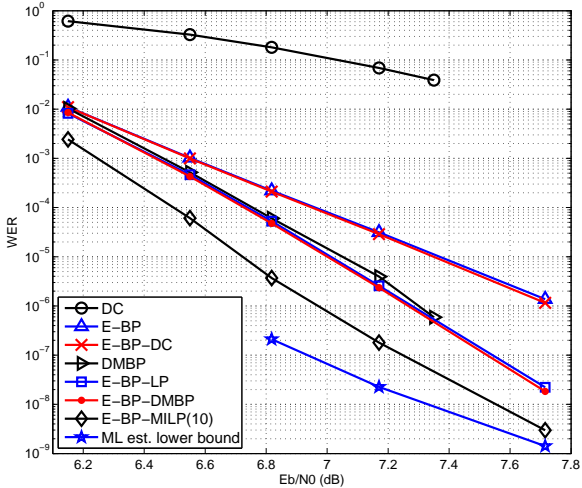} 
      \label{N2209BSC_a}} 
\quad 
\subfigure[Results when $T_{\max} =
    300$
    iterations]{\includegraphics[width=3.5in]{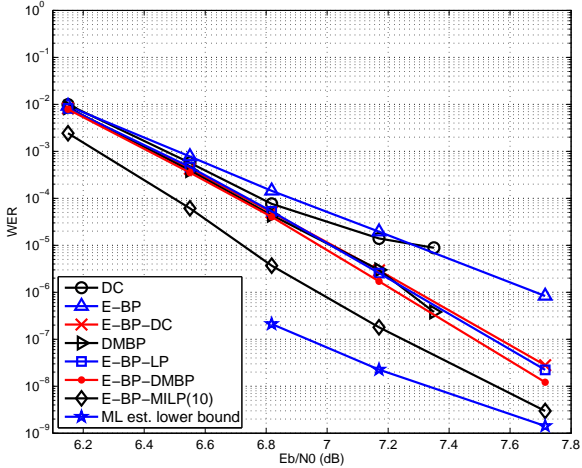}
      \label{N2209BSC_b}}
\end{center}
  \caption {Error performance comparisons for a length-2209,
    rate-0.916 array LDPC code over the BSC.} \label{N2209BSC}
\end{figure}

Figure \ref{N2209BSC} depicts the WER performance comparison of the
length-2209 array LDPC code over the BSC. For this QC-LDPC code, we
observe broadly similar performance to the random LDPC code.


\begin{figure}[t]
\begin{center}
\subfigure[Results when $T_{\max} = 50$
    iterations]{\includegraphics[width=3.5in]{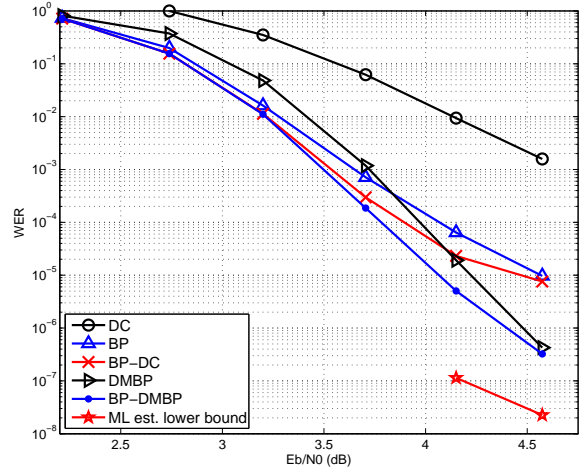}
      \label{N1057AWGN_a}} 
\quad 
\subfigure[Results when $T_{\max} =
    200$
    iterations]{\includegraphics[width=3.5in]{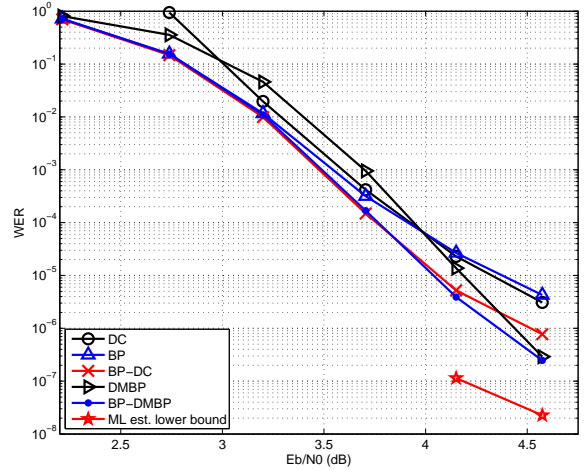}
      \label{N1057AWGN_b}}
\end{center}
  \caption {Error performance comparisons for a length-1057, rate-0.77
    random LDPC code over the AWGN channel.} \label{N1057AWGN}
\end{figure}

Figure \ref{N1057AWGN} shows the WER performance comparison of the
length-1057 random LDPC code over the AWGN channel. We observe that
the BP decoder for this code exhibits an error floor. DMBP 
improves the error floor performance compared with BP and
does not have an apparent error floor. When 200 iterations are used,
the DC decoder
has a similar performance to BP. In the high SNR region, the DC decoder does not converge
to an incorrect codeword as frequently as it does over the BSC. Note also that
on the AWGN channel, while the DMBP decoder outperforms BP in the
error-floor regime, it actually starts out worse in the low SNR
regime.


\begin{figure}[t]
\begin{center}
\subfigure[Results when $T_{\max} = 50$
    iterations]{\includegraphics[width=3.5in]{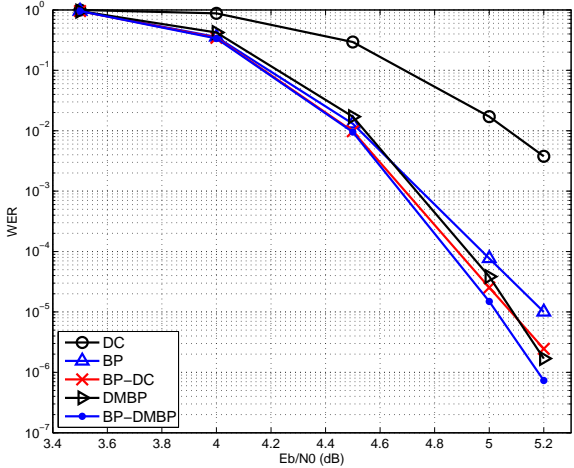} 
      \label{N2209AWGN_a}} 
\quad 
\subfigure[Results when $T_{\max} =
    200$ or $500$
    iterations]{\includegraphics[width=3.5in]{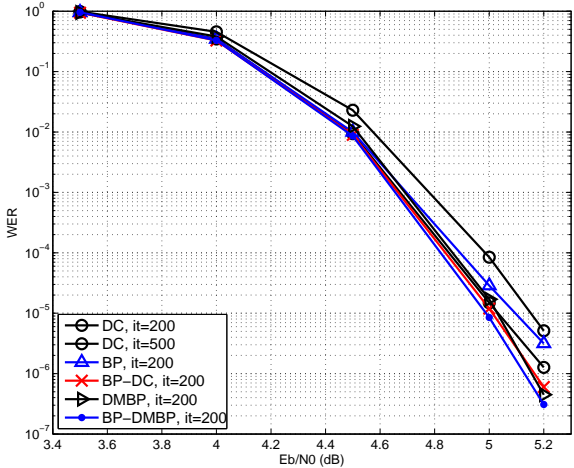}
      \label{N2209AWGN_b}}
\end{center}
  \caption {Error performance comparisons for a length-2209 and
    rate-0.916 array LDPC code over the AWGN
    channel.} \label{N2209AWGN}
\end{figure}

Figure \ref{N2209AWGN} depicts the WER performance comparison of the
length-2209 array LDPC code over the AWGN channel. For this QC-LDPC
code, we observe similar performance to the random LDPC code. Note again that while
all decoders benefit from additional allowed iterations, the
DC decoder in particular becomes increasingly competitive as the number of allowed
iterations increases.


Our basic motivation for the DC and DMBP decoders was that the
difference-map dynamics may help a decoder avoid dynamical ``traps''
that could be related to the trapping sets that are believed to cause
error floors. The very good performance of the DMBP decoder in the error
floor regime indicates that there may in fact be
a reduction in the number of trapping sets,
but on the other hand, some trapping sets clearly continue
to exist, even for the DMBP decoder.
In particular, we followed the approach of \cite{Dolecek} and
performed some preliminary investigations of individual
``absorbing sets'' in the array code that they studied,
and found that although the DMBP decoder performed better on average than
the BP decoder,
it still would not escape if started sufficiently close to 
particular difficult absorbing sets.

\section{Conclusion}
In this paper, we investigate two decoders for LDPC codes: a DC
decoder that directly applies the divide and concur approach to decoding LDPC
codes, and a DMBP decoder that imports the difference-map idea
into a min-sum BP-type decoder. The DMBP decoder shows
particularly promising improvements in error-floor performance
compared with the standard sum-product BP decoder, with comparable
computational complexity, and is amenable to hardware implementation.

The DMBP decoder can be criticized for lacking a solid theoretical basis: it
was constructed using intuitive ideas and is mostly interesting because of its
excellent performance. The fact that its performance closely parallels that
of linear programming decoders suggests that it might be related to them. 
In fact, our work was partially motivated by our earlier results which showed that
LP decoders can significantly improve upon BP performance in the error floor
regime \cite{Multistage}; 
we aimed to develop a message-passing decoder that could reproduce LP performance
with complexity similar to BP. 

Work in the direction of creating an efficient
message-passing linear programming decoder that could replace LP solvers that relied
on simplex or interior point methods was begun
by Vontobel and Koetter \cite{VKLP}, and message-passing algorithms that
converge to an LP solution for some problems 
were suggested by Globerson and Jaakkola \cite{Globerson}.
Our DMBP update equations are quite similar to those in the
GEMPLP algorithm suggested by Globerson and Jaakkola, but our limited experiments
with a GEMPLP decoder show that it does not reproduce LP decoding performance. For that
matter, we have been unable to devise any other message-passing decoder with complexity
similar to BP
that exactly reproduces linear programming decoding.
Elucidating the
precise relationship
between DMBP and LP decoders remains an outstanding theoretical problem, but from the
practical point of view, our results show that the DMBP decoder already serves as
an efficient message-passing decoder that significantly improves error floor performance
compared with standard BP.

\bibliographystyle{IEEE}

\begin{thebibliography}{10}

\bibitem{ModernCodingTheory} T. Richardson and R. Urbanke, {\em Modern
    Coding Theory}. Cambridge University Press, 2008.

\bibitem{FreyKoetterVardy} B.~J.~Frey, R. Koetter, and A. Vardy, ``Signal-space
characterization of iterative decoding'', {\em IEEE Trans. Inform. Theory}, vol. 47,
Feb. 2001, pp. 766-781.

\bibitem{GraphCover} P.~O.~Vontobel and R.~Koetter, ``Graph-cover decoding and
finite-length analysis of message-passing iterative decoding of LDPC codes,''
to appear in {\em IEEE Trans. Inform. Theory} http://www.arxiv.org/abs/cs.IT/0512078.

\bibitem{NearCodeword} D. MacKay and M. Postol, ``Weaknesses of
  Margulis and Ramanujan-Margulis low-density parity-check codes,''
  {\em Electronic Notes in Theoretical Computer Science}, vol. 74,
  2003.

\bibitem{trappingSet} T. Richardson, ``Error floors of LDPC codes,''
  {\em Proc. 41st Allerton Conf. on Communications, Control, and
    Computing}, Allerton House, Monticello, IL, Oct. 2003.

\bibitem{Dolecek} L. Dolecek, P. Lee, Z. Zhang, V. Anatharam,
  B. Nikolic, and M.J. Wainwright, ``Predicting error floors of
  structured LDPC codes: deterministic bounds and estimates,'' to
  appear in {\em IEEE Jour. Sel. Areas in Comm.}, 2009.

\bibitem{VontobelWebpage} See
  www.hpl.hp.com/personal/Pascal\_Vontobel/pseudocodewords/papers/
  for a collection of papers on pseudocodewords and related ideas.

\bibitem{PEG} X.-Y. Hu, E. Eleftheriou, and D.~M.~Arnold. ``Regular and
  irregular progressive edge-growth Tanner graphs,'' {\em IEEE
    Trans. Inform. Theory}, vol. 51, no. 1, Jan. 2005, pp. 386--398.

\bibitem{ACE} T. Tian, C. Jones, J. D. Villasenor, and R. D. Wesel,
  ``Construction of irregular LDPC codes with low error floors,'' {\em
    IEEE International Conference on Communications}, vol. 5,
  Anchorage, AK, May 2003, pp. 3125--3129.

\bibitem{HighGirth} Y. Wang, J.~S.~Yedidia, S.~C.~Draper, ``Construction
  of high-girth QC-LDPC codes,'' {\em Fifth International Symposium on
    Turbo Codes and Related Topics} 2008.  Available online at
  http://www.merl.com/publications/TR2008-061/.

\bibitem{EGLDPC} J. Zhang, J.~S.~Yedidia, M.~P.~C.~Fossorier,
  ``Low-latency decoding of EG LDPC codes,'' {\em Journal of Lightwave
    Technology}, vol. 25, Sept. 2007, pp. 2879--2886.

\bibitem{GLDPC} R. M. Tanner, ``A recursive approach to low complexity
  codes,'' {\em IEEE Trans. Inform. Theory}, vol. 27, Sept. 1981,
  pp. 533--547.

\bibitem{DGLDPC} Y. Wang and M. Fossorier, ``Doubly generalized LDPC
  codes,'' {\em IEEE Int. Symp. on Inform. Theory}, Seattle, WA,
  Jul. 2006. pp. 669--673.

\bibitem{gallager} R. G. Gallager, {\em Low-Density Parity-Check
    Codes}. Cambridge, MA: M.I.T. Press, 1963.


\bibitem{HanRyan} Y. Han and W. E. Ryan, ``Low-floor decoders for LDPC
  codes,'' {\em Forty-Fifth Annual Allerton Conference}, Allerton, IL,
  Sept. 2007, pp. 473--479.

\bibitem{MIALP} S. C. Draper, J. S. Yedidia, and Y. Wang, ``ML
  decoding via mixed-integer adaptive linear programming,'' {\em
    Proc. IEEE Int. Symp. Information Theory}, Nice, France,
  Jun. 2007, pp. 1656--1660.

\bibitem{Multistage} Y. Wang, J. S. Yedidia, and S. C. Draper ,
  ``Multi-stage decoding of LDPC codes,'' {\em Proc. IEEE
    Int. Symp. Information Theory}, Seoul, Korea,
  Jun. 2009. pp. 2151--2155.

\bibitem{DAC} S. Gravel and V. Elser, ``Divide and concur: a general
  approach to constraint satisfaction,'' {\em Phys. Rev.} E 78,
  2008. p. 036706. Available online at {\em arXiv:0801.0222v1}.

\bibitem{mezardMontanari09} M.~M\'ezard and A.~Montanari, {\em
    Information, Physics, and Computation}, Oxford University Press,
  2009. Chapter 8.


\bibitem{GravelThesis} S. Gravel, {\em Using Symmetries to Solve Asymmetric Problems},
Ph.D. Thesis, Cornell University 2009. See section 5.4.

\bibitem{Fienup} J.~R.~Fienup, ``Phase retrieval algorithms: a
  comparison,'' {\em Applied Optics}, vol. 21, 1982, pp. 2758--2769.

\bibitem{ElserPNAS} V.~Elser, I.~Rankenburg, and P.~Thibault,
  ``Searching with iterated maps,'' {\em Proc. Nat. Acad. Sci. USA},
  vol. 104, 2007, pp. 418--423.

\bibitem{Bauschke} H.~H.~Baushke, P.~L.~Combettes, and D.~R.~Luke, ``Phase retrieval,
error reduction algorithm, and Fienup variants: a view from convex optimization,''
{\em J. Op. Soc. Am. A}, vol. 19, July 2002, pp. 1334-1345.

\bibitem{FactorGraph} F.~R.~Kschischang, B.~J.~Frey, and
  H.-A.~Loeliger, ``Factor graphs and the sum-product algorithm,''
  {\em IEEE Trans. on Info. Theory}, vol. 47, Feb. 2001, pp. 498--519.

\bibitem{DataBase} ``Encyclopedia of Sparse Graph Codes,'', available
  online at
  http://www.inference.phy.cam.ac.uk/mackay/codes/data.html. Code
  1057.244.3.457.

\bibitem{Fan} J. L. Fan, ``Array codes as low-density parity-check
  codes,'' {\em Proc. 2nd Int. Symp. Turbo Codes}, Brest, France,
  Sept. 2000, pp. 545--546.

\bibitem{richardsonUrbanke:01} T.~J.~Richardson and R.~Urbanke, ``The
  capacity of low-density parity-check codes under message-passing
  decoding,'' {\em IEEE Trans. on Info. Theory}, vol. 47, Feb. 2001,
  pp. 599--618.

\bibitem{Taghavi} M.-H.~N.~Taghavi and P.~H.~Siegel, ``Adaptive methods for
linear programming decoding,'' {\em IEEE Trans. Inform. Theory}, vol. 54, Dec. 2008,
pp. 5396-5410.

\bibitem{GLPK} ``GNU Linear Programming Kit,'' http://www.gnu.org/software/glpk.

\bibitem{VKLP} P.~Vontobel and R.~Koetter, 
``Towards low-complexity linear-programming decoding,'' {\em
Proc. 4th Int. Symposium on Turbo Codes and Related Topics}, Munich, Germany, 2006.

\bibitem{Globerson} A. Globerson and T.~Jaakkola, 
``Fixing max-product: convergent message passing algorithms
for MAP LP-relaxations,'' {\em Advances in Neural Information Processing Systems} 20,
Vancouver, Canada, 2007.

\end{thebibliography}

%

%

\end{document}